\documentclass[aps,prb,twocolumn]{revtex4}
\usepackage{amssymb}
\usepackage{graphicx}

\begin{document}
\title{Superconductivity above 10~K in Non-Cuprate Oxides}
\author{David C. Johnston}
\affiliation{Ames Laboratory and Department of Physics and Astronomy, Iowa State University, Ames, Iowa 50011, USA}
\date{\today}
\begin{abstract}
Beginning in 1973, several non-cuprate transition metal and non-transition metal oxides were discovered with superconducting transition temperatures between 10 and 30~K\@.  Retrospectives about these discoveries are given.\\

Invited paper submitted for publication in a book entitled \emph{100 Years of Superconductivity}, edited by H.\ Rogalla and P.\ Kes.

\end{abstract}
\maketitle

\section{Introduction}

The quest for high superconducting transition temperatures $T_{\rm c}$ led to a slow increase in the maximum observed $T_{\rm c}$ from 4.2~K in Hg in 1911, which was the discovery of superconductivity itself by Onnes, to 22--23~K in thin films of the cubic A-15 structure compound ${\rm Nb_3Ge}$ as reported by Gavaler et al.\ in 1973 and 1974.\cite{Gavaler1973}  Oxides and non-transition metal compounds were not seriously considered by most researchers as contenders for high $T_{\rm c}$.  Superconducting oxides such as NbO and doped ${\rm SrTiO_3}$ exhibited $T_{\rm c}\lesssim 1$~K\@.  The highest $T_{\rm c}$ oxide, hexagonal rubidium tungsten bronze Rb$_x$WO$_3$, showed $T_{\rm c}\leq 6.6$~K.\cite{Remeika1967}  Until 1986, ``high $T_{\rm c}$'' was considered to be any $T_{\rm c}$ above 10~K\@.

Beginning in 1973, a series of discoveries of $T_{\rm c} > 10$~K in oxides and non-transition metal compounds was made that changed our view of their potential for high $T_{\rm c}$.  A maximum $T_{\rm c}$ onset of 13.7~K was discovered by the author and coworkers in 1973 for ${\rm LiTi_2O_4}$ with the cubic spinel structure,\cite{Johnston1973} and $T_{\rm c}$ up to about 13~K was found by Sleight et al.\ for the non-transition metal oxide Ba(Pb$_{1-x}$Bi$_x$)O$_3$ with the cubic perovskite structure in 1975.\cite{Sleight1975}  These $T_{\rm c}$s were the highest for oxide and/or non-transition metal compounds until the discovery by Bednorz and M\"uller in 1986 of superconductivity up to about 30~K in the layered copper oxide compound La$_{2-x}$Ba$_x$CuO$_4$ that contains Cu square lattice layers,\cite{Bednorz1986, Johnston1997} as discussed in other chapters by Alex M\"uller and by Paul Chu.  The current maximum $T_{\rm c}$ of 164~K for this class of compounds was obtained under pressure by Gao et al.\ in 1994 and is currently also the record $T_{\rm c}$ for any material.\cite{Gao1994}  This $T_{\rm c}$ was reached by applying 31~GPa pressure to a Hg$_{1-x}$Pb$_x{\rm Ba_2Ca_2Cu_3}$O$_{8+\delta}$ sample that had a zero-pressure $T_{\rm c}$ of 134~K.\cite{Gao1994}  Superconductivity at temperatures up to about 30~K was found in the non-transition-metal cubic perovskite oxide compound (Ba$_{1-x}$K$_x$)BiO$_3$ by Mattheiss et al.\cite{Mattheiss1988} and  Cava et al.\cite{Cava1988} in 1988.  In related developments, bulk high $T_{\rm c}$s were discovered by Tanigaki et al.\ in non-transition-metal alkali metal-doped $A_x$C$_{60}$ Buckyballs up to about 33~K in 1991,\cite{Tanigaki1991} and in MgB$_2$ at 39~K by Nagamatsu et al.\cite{Nagamatsu2001} in 2001 as reviewed in a separate chapter by Jun Akimitsu and Takahiro Muranaka.  In 2008, the Fe-containing tetragonal compound LaFeAsO$_{1-x}$F$_x$ was discovered by Kamihara et al.\ to have a high $T_{\rm c} = 26$~K,\cite{Kamihara2008} as discussed by Hideo Hosono in another chapter.  Members of this general class of materials have  crystal structures containing iron square lattice layers with a maximum $T_{\rm c}$ of 56~K,\cite{Johnston2010} which coincidentally(?) is the same transition metal sublattice structure as the Cu atoms have in the layered cuprate high-$T_{\rm c}$ superconductors.

Herein are presented retrospectives of the discoveries of $T_{\rm c} > 10$~K in the non-cuprate oxides ${\rm LiTi_2O_4}$, Ba(Pb$_{1-x}$Bi$_x$)O$_3$, and (Ba$_{1-x}$K$_x$)BiO$_3$, with an emphasis on ${\rm LiTi_2O_4}$ because of the author's familiarity with this compound.  Lack of space precludes discussions of their normal or superconducting state physical properties other than their $T_{\rm c}$s and of their mechanism(s) for superconductivity.

\section{${\rm LiTi_2O_4}$}

\begin{figure}
\includegraphics [width=1.9in]{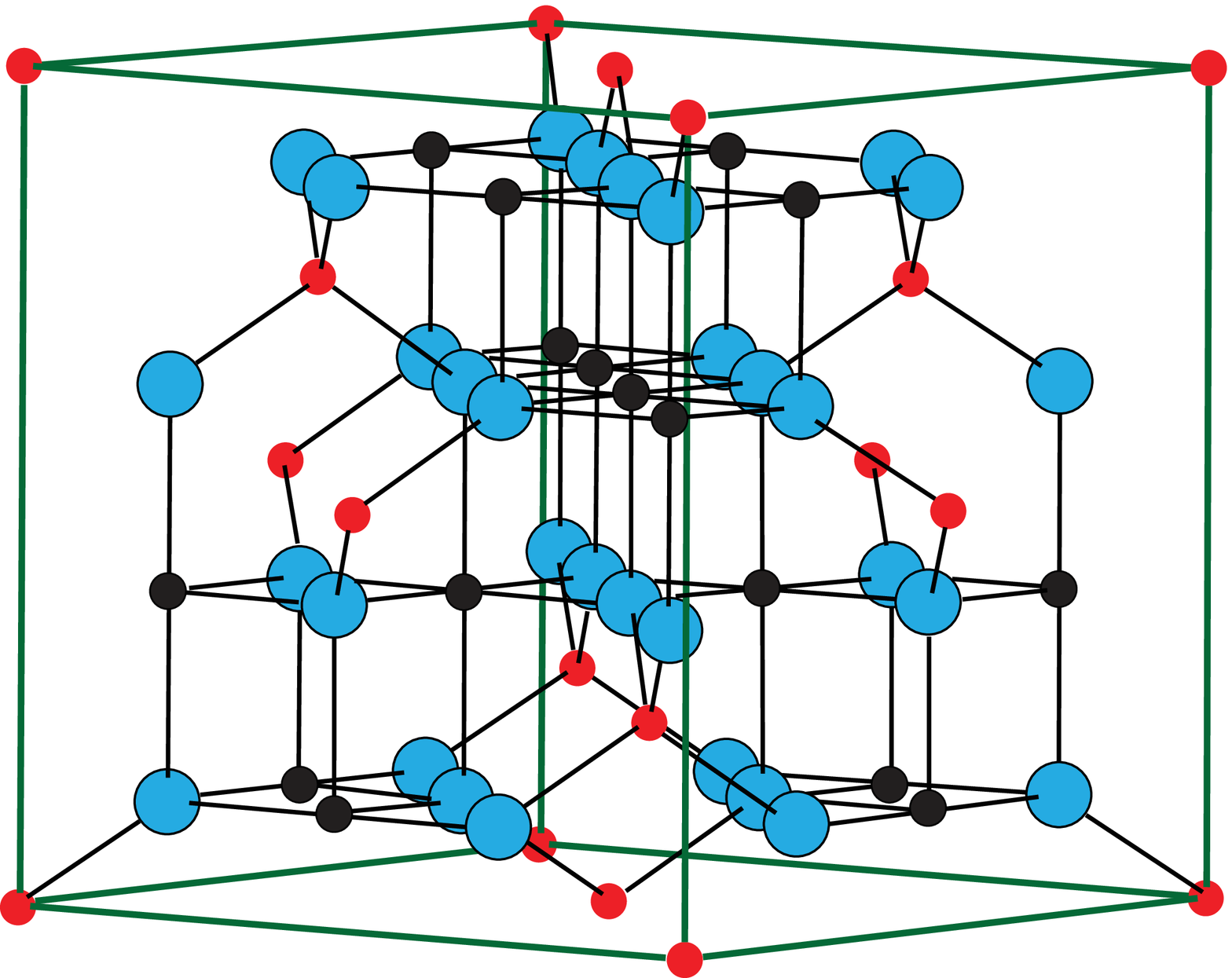}\vspace{0.2in}
\includegraphics [width=1.9in]{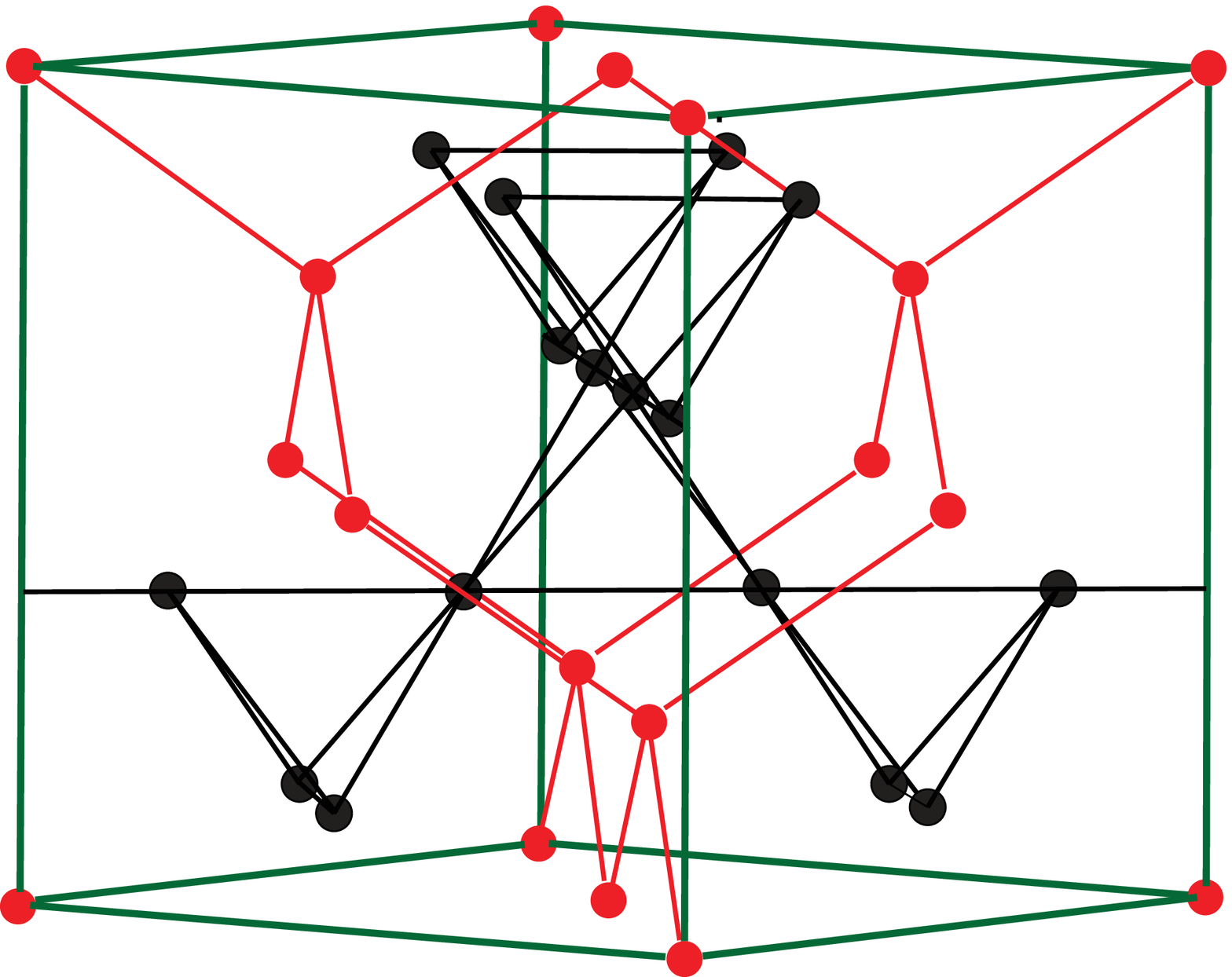}\vspace{0.2in}
\includegraphics [width=1.6in]{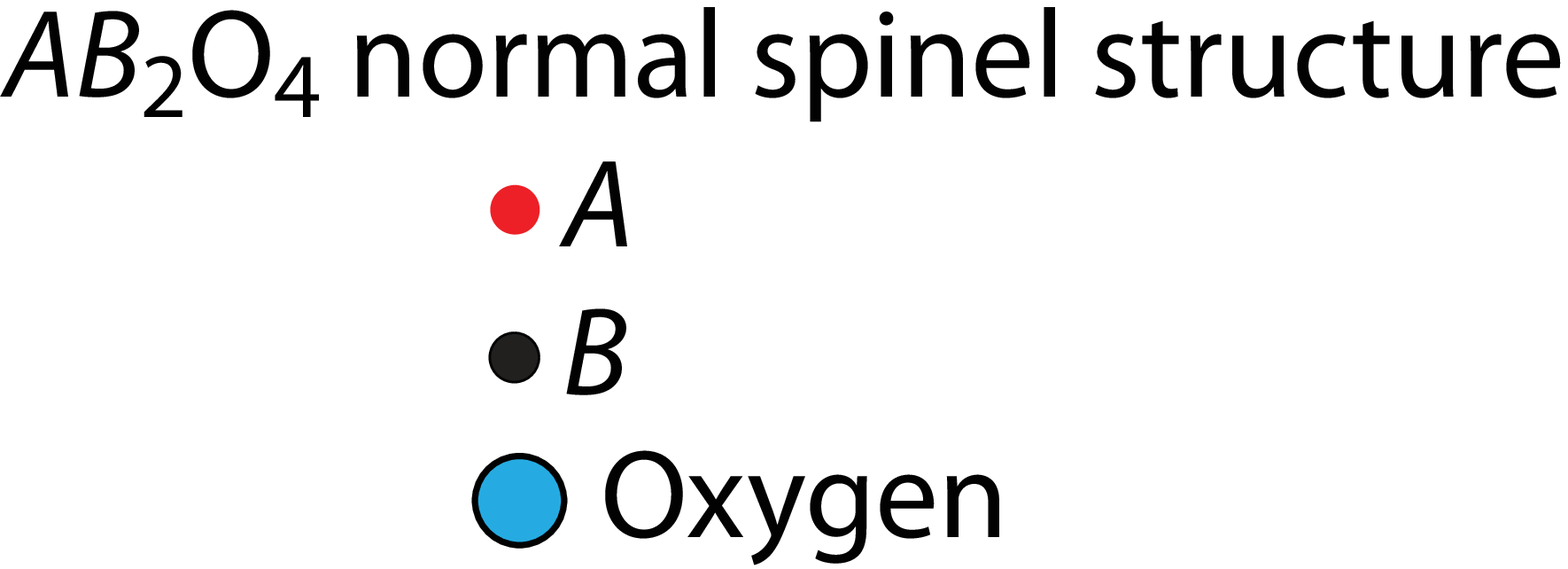}
\caption{(Color online) (top panel) Face-centered-cubic crystal structure of a normal spinel oxide compound in which the $A$ atoms occupy only tetrahedral sites and the $B$ atoms only the octrahedral sites between O layers that are close-packed along the [111] directions.  The unit cell edges are shown.  The unit cell contains eight formula units.  (bottom panel) Cation sublattice of the normal spinel structure.  The $B$ sublattice consists of corner-sharing tetrahedra.}
\label{StructSpinel} 
\end{figure}

The face-centered-cubic normal spinel crystal structure of ${\rm LiTi_2O_4}$ is shown in Fig.~\ref{StructSpinel} (space group \emph{Fd}$\bar{3}$\emph{m}, No.~227).  The structure prototype is the mineral spinel with the composition ${\rm MgAl_2O_4}$.  The structure consists of a nearly cubic-close-packed array of O atoms layered along the [111] directions of the unit cell, with the Mg atoms in tetrahedral interstices and Al atoms in octahedral interstices between adjacent O layers.  The cation sublattices, shown separately in the bottom panel of Fig.~\ref{StructSpinel}, are both geometrically frustrated for antiferromagnetic ordering due to the triangular connectivity of the respective nearest neighbors.  Most 3$d$ transition metal spinel structure compounds become distorted at low temperatures, which can partially relieve the frustration.  There are very few transition metal oxide spinel compounds that remain metallic  and cubic at low temperatures.  The intermediate-valent spinel compound ${\rm LiTi_2O_4}$ is one of these and ${\rm LiV_2O_4}$ is another, where the formal oxidation state of both the Ti ($d^{0.5}$) and V ($d^{1.5}$) is +3.5.  This nonintegral valence state, together with the undistorted crystal structure, require both compounds to be metals to low temperatures.

The discovery of superconductivity in ${\rm LiTi_2O_4}$ came about in the following way.  At about the middle of the period (1969--1975) when I was a graduate student at the University of California at San Diego, my research advisor, Bernd T. Matthias, asked me to make a sample of Li$_x{\rm Ti_{1.1}S_2}$, which he and coworkers at Bell Laboratories, Murray Hill, NJ, had recently (1972) reported to become superconducting at a high temperature of 10--13~K for $0.1 < x \leq 0.3$.\cite{Barz1972}  Lithium is a very reactive alkali metal, so I asked a chemistry professor to make some powder of Li$_x{\rm Ti_{1.1}S_2}$ that I could process further.  His postdoc Hari Prakash, using wet chemical methods, provided me with a batch of amorphous Li$_x{\rm Ti_{1.1}S_2}$ which, in retrospect, probably had a high surface area.  In order to make a crystalline sample, I wanted to melt it in an arc-furnace (like a commercial arc-welder, but in an inert argon atmosphere).  This was certainly not an optimum way to crystallize the material!  However, even so, I wanted to do this experiment, so I first poured some of the powder out of the bottle in air in order to make a pressed pellet, but the powder started to smoke!  Therefore I made the pellet as fast as possible and loaded it into the arc-furnace.  When I tried to melt the pellet, the sample made a lot of smoke that filled up the arc-furnace because of the volatility of Li and S, but it also made a melted round ball that was a beautiful dark blue color (metals are usually silvery and don't usually have colors; pure bulk elemental Os metal is light blue, though).  Then I measured pieces of the blue ball for superconductivity using ac susceptibility measurements and found that the sample became superconducting at 11~K\@.  This seemed to confirm Bernd's findings.  But all was not as it seemed.  My powder x-ray diffraction pattern of the sample contained lines that I could index as ${\rm Li_2TiO_3}$, and I subsequently found that the intensity of these lines correlated with the volume fraction of superconductivity in similarly prepared samples.

Then, as luck would have it, Bernd had a famous crystallographer friend, Willie Zachariasen, who happened to be visiting Bernd's lab at UCSD at the time.  I pointed out to him the set of x-ray diffraction peaks that correlated with the volume fraction of superconductivity.  After a couple of days analyzing the peaks, he figured out that the composition of the superconducting material in my samples was close to ${\rm LiTi_2O_4}$ with a face-centered-cubic (fcc) normal-spinel structure.  It turned out that the strongest lines in the pattern were at about the same positions as those for the different phase ${\rm Li_2TiO_3}$ noted above.  Undoubtedly Hari did provide me with pure Li$_x{\rm Ti_{1.1}S_2}$, but when I exposed it to air, it reacted with the air and oxidized before I could make it into a pellet.  Then when I subsequently arc-melted it in pure Ar, part of the sample turned into ${\rm LiTi_2O_4}$ which was superconducting at 11~K\@.  Furthermore, I later discovered that the metallic spinel structure compound ${\rm LiTi_2O_4}$ transforms to a semiconducting Ramsdellite structure compound with the same composition upon heating above 950~$^\circ$C, and that the conversion of the structure to the spinel structure on cooling to below 950~$^\circ$C is a slow process.  Apparently when the melted sample cooled in the arc-furnace, at least some of it converted to the low-temperature spinel crystal structure.  The discovery of superconductivity in ${\rm LiTi_2O_4}$ was thus pure serendipity. 

I subsequently learned how to synthesize ${\rm LiTi_2O_4}$ in a rational way and further confirmed that the superconductivity arose from ${\rm LiTi_2O_4}$ with the spinel structure at temperatures up to 13.7~K\@.  R.~Viswanathan at UCSD carried out heat capacity measurements on the initial samples and demonstrated that ${\rm LiTi_2O_4}$ is a bulk superconductor.  The first announcement of ``High-Temperature Superconductivity in the Li-Ti-O Ternary System" was in 1973.\cite{Johnston1973}  I did my Ph.D. Thesis on the follow-up synthesis, structure, and properties of polycrystalline samples of ${\rm LiTi_2O_4}$ and of the solid solution Li$_{1+x}$Ti$_{2-x}$O$_4$.\cite{Johnston1976, McCallum1976, Shelton1976}  Some of these studies were carried out in collaboration with Bill McCallum, Carlos Luengo, Brian Maple, Robert Shelton and Hermann Adrian who were all also at UCSD at the time.

Before I worked on ${\rm LiTi_2O_4}$, it was known that there exists a complete solid solution of spinel-structure compounds from ${\rm LiTi_2O_4}$ to ${\rm Li[Li_{1/3}Ti_{5/3}]O_4}$, which can be written Li$_{1+x}$Ti$_{2-x}$O$_4$ with $0 \leq x \leq 1/3$.\cite{Johnston1976}  In this solid solution, some Li substitutes for the Ti on the octahedral sites of the structure.  At the same time, the formal oxidation state of the Ti changes from +3.5 to +4.0.  The compound with $x = 1/3$ is a colorless nonmagnetic insulator.  Therefore there had to be a metal-to-insulator transition with increasing $x$.  As part of my Thesis work, I investigated this issue and found a metal-insulator transition at $x \approx 0.10$ with $T_{\rm c}$ being nearly constant for $0 \leq x \leq 0.1$.  This suggested that if a metal-insulator transition occurred in a specific materials system, the potential for high $T_{\rm c}$ superconductivity might be greatest for the metallic compound in that system with a composition closest to the metal-insulator boundary.  I made a compilation of oxide and chalcogenide compounds and found that this correlation indeed held about 80\% of the time.  It was subsequently found to be applicable to the cuprate high-$T_{\rm c}$ superconductors.  This correlation is also mentioned in Arthur Sleight's commentary below.  A summary of the literature on ${\rm LiTi_2O_4}$ up to 1999 was given by Moshopoulou.\cite{Moshopoulou1999}

Before leaving the topic of ${\rm LiTi_2O_4}$, one might think that the very similar metallic normal spinel structure compound ${\rm LiV_2O_4}$, which has the same beautiful dark blue color, would also be a superconductor.  However, in 1973, it was already known that ${\rm LiV_2O_4}$ was not superconducting above 4~K, and that this compound, despite being metallic, showed a Curie-Weiss temperature dependence to the magnetic susceptibility with a Curie constant indicating that each V atom had a spin $S = 1/2$ with a spectroscopic splitting factor $g \sim 2$.  Since my advisor Bernd Matthias was mainly interested in discovering new superconductors at UCSD, I did not pursue the properties of ${\rm LiV_2O_4}$ at that time.  It was not until 1997 that my group at Iowa State University discovered that ${\rm LiV_2O_4}$ is a very rare example of a $d$-electron heavy fermion compound, in collaboration with Clayton Swenson and Ferdinando Borsa's and Alan Goldman's groups at Iowa State and with the groups of James Jorgensen, Brian Maple and Yasutomo Uemura.\cite{Kondo1997, Johnston2000}

\section{B\lowercase{a}(P\lowercase{b}$_{1-x}$B\lowercase{i}$_x$)O$_3$}

\begin{figure}
\includegraphics [width=1.6in]{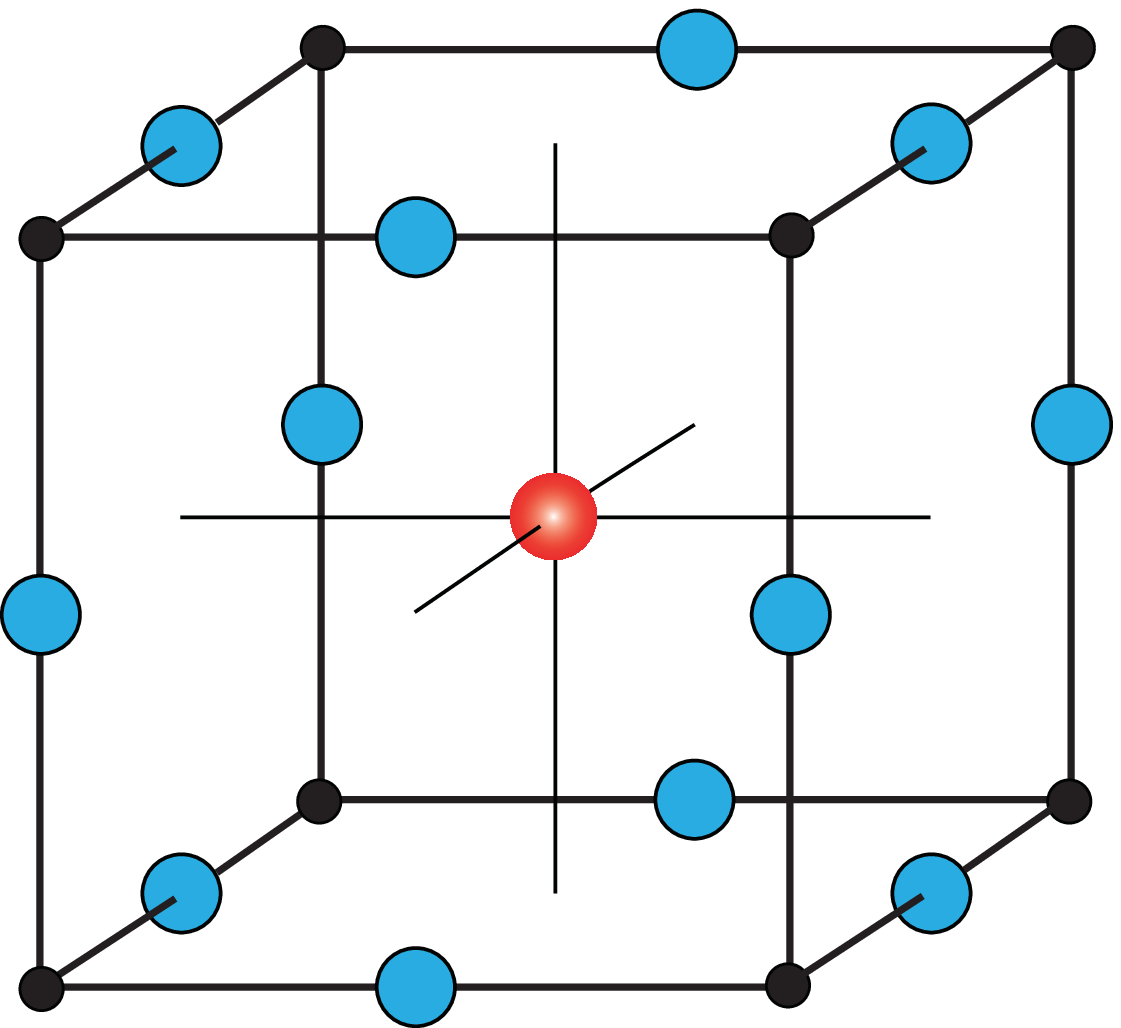}\vspace{0.2in}
\includegraphics [width=1.8in]{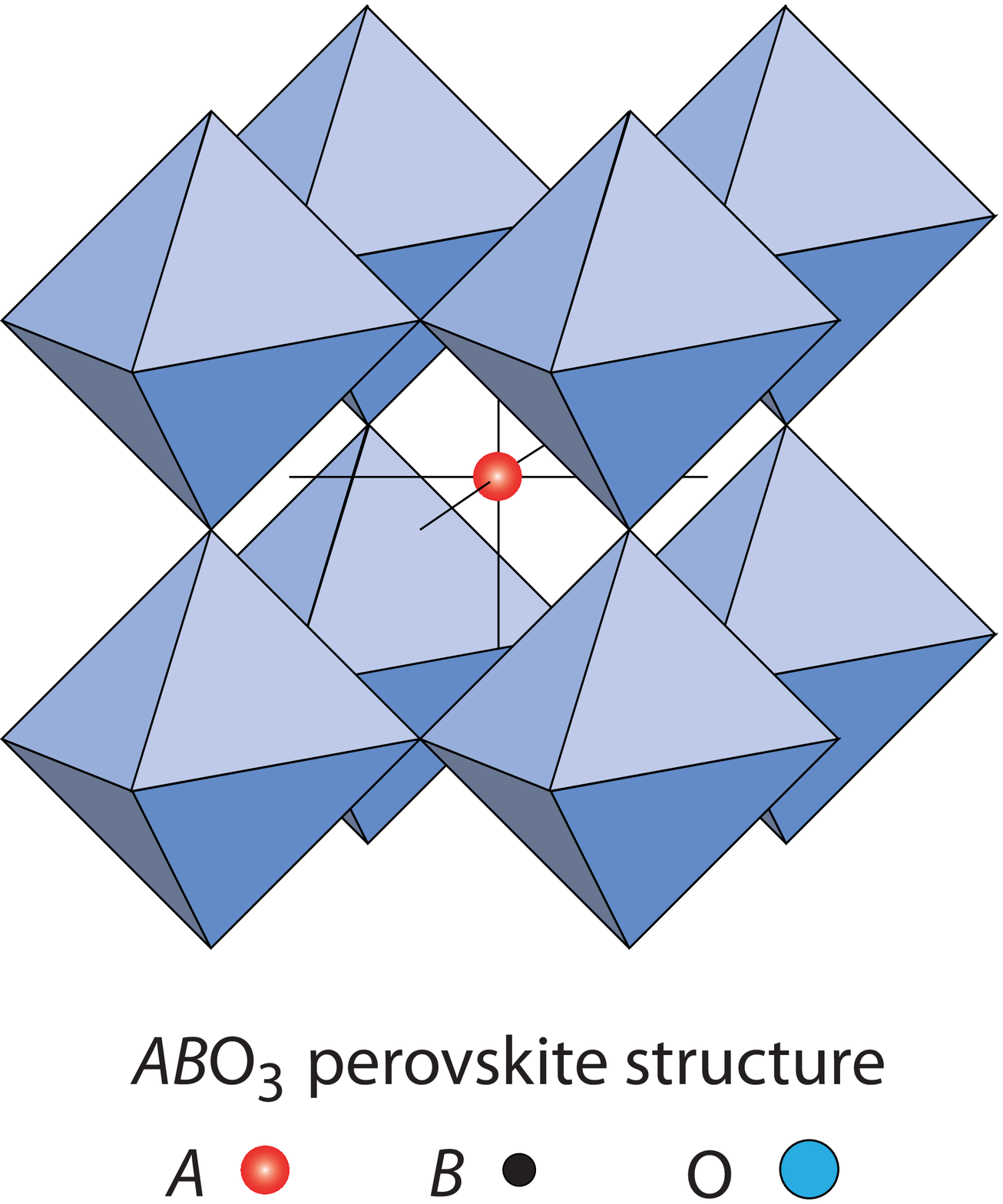}
\caption{(Color online) (top panel) Cubic perovskite crystal structure of an $AB$O$_3$ compound.  The unit cell edges are shown.  The unit cell contains one formula unit.  (bottom panel)  Figure showing the corner-sharing $B$O$_6$ octahedra with the $A$ atoms in between.  The centers of the eight $B$O$_6$ octrahedra shown are each occupied by $B$ atoms and are at the corners of the unit cell shown in the top panel.}
\label{StructPerovskite} 
\end{figure}

This compound has a primitive cubic perovskite crystal structure (space group \emph{Pm}$\bar{3}$\emph{m}, No.~221), named after the prototype mineral perovskite, CaTiO$_3$.  The structure is shown in Fig.~\ref{StructPerovskite}.  The structure of an \emph{AB}O$_3$ cubic perovskite compound consists of corner-sharing \emph{B}O$_6$ octahedra with $A$ atoms in the tunnels between the octahedra.  The structure is very susceptible to lattice distortions, sometimes by cooling the cubic structure from room temperature, and many insulating compounds with distorted variants of this structure are ferroelectric.

The following account of the discovery of superconductivity in the Ba(Pb$_{1-x}$Bi$_x$)O$_3$ system at temperatures up to 13~K in 1975 was provided by Arthur Sleight, showing that serendipity was again at work.

``I had just discovered a first-order metal-insulator [MI] transition in ${\rm Tl_2Ru_2O_7}$ and a second-order metal-insulator transition in ${\rm Cd_2Os_2O_7}$.  I was wondering whether there could be a metal-insulator transition in an oxide without a transition metal.  BaBiO$_3$ was known, but it had not been well characterized.  The diffraction data that existed at that time suggested that there might be only one type of Bi present.  If this were true, BaBiO$_3$ should be metallic with a half-filled 6$s$ conduction band.  I was thinking that as a function of temperature there might be a transition from BaBi$^{4+}$O$_3$ (metal) to Ba$_2$Bi$^{3+}$Bi$^{5+}$O$_6$ (insulator).  I decided crystals were needed, and none have ever been reported.  I grew some crystals.  At first, I thought they might be metallic because they were a gold color with a metallic luster.  However, electrical measurements showed that the crystals were semiconducting with a rather high resistivity.  I knew that BaPbO$_3$ was a very good conductor.  So I thought there must be some sort of metal-insulator transition in the BaPbO$_3$--BaBiO$_3$ solid solution, which had never been investigated.  I first made 50/50 (${\rm BaPb_{1/2}Bi_{1/2}O_3}$).  It was not metallic. So I then made ${\rm BaPb_{0.75}Bi_{0.25}O_3}$, and it was metallic and superconducting.  \ldots I wanted to be very sure of things before we published.  I grew crystals of ${\rm BaPb_{0.75}Bi_{0.25}O_3}$ and they were superconducting.  We were not set up for magnetic measurements appropriate for superconductors.  But I gave a sample to a guy who measured magnetic susceptibility.  He said it must be a superconductor because the sample jumped out of its holder when the temperature of zero resistivity was reached.  I was confident of our results, and I did not want our paper to be reviewed by skeptics.  So I sent it to a journal where an Editor would communicate it.\cite{Sleight1975}  I then tried BaBiO$_3$--KBiO$_3$ at that time, but only once and saw no superconductivity.''

``I had previously worked on W and Re oxides superconductors, and I did appreciate at that time that superconductivity frequently occurred in compositions that were close to a phase boundary between metallic and insulating properties.  The actual value of $T_{\rm c}$ in the Ba(Bi,Pb)O$_3$ system was a surprise to me.   So I was not really looking for a new superconductor, but my search for a MI transition based on BaBiO$_3$ made this discovery inevitable.''

\section{(B\lowercase{a}$_{1-x}$K$_x$)B\lowercase{i}O$_3$}

Superconductivity was discovered in 1988 at temperatures up to around 22~K in the system (Ba$_{1-x}$K$_x$)BiO$_3$ by Mattheiss et al.\cite{Mattheiss1988} using a specific focused approach.  Several factors led to this discovery.\cite{Mattheiss1988}  First, the superconductivity at temperatures up to 13~K in the related perovskite system Ba(Pb$_{1-x}$Bi$_x$)O$_3$ described above was well-known.  Second, according to Mattheiss, et al., based on previous work, ``These results suggest that it should be possible to suppress the ordering waves [that result in insulating behavior] and extend the metallic regime in BaPb$_{1-x}$Bi$_x$O$_3$ closer to the half-filled band condition (BaBiO$_3$) where the electron-phonon interaction is a maximum, by leaving the conducting Bi-O complex intact and instead doping substitutionally at the inactive Ba donor sites.  This situation, which is analogous to that in La$_{2-x}$(Ba,Sr)$_x$CuO$_4$, is expected to produce marginal stability and enhanced $T_{\rm c}$'s.  In earlier studies, a combination of K and Pb doping in a ${\rm Ba_{0.9}K_{0.1}Pb_{0.75}Bi_{0.25}O_3}$ sample has produced similar critical temperatures ($T_{\rm c} \sim 12$~K) but sharper transitions than those observed in Pb-doped ${\rm BaPb_{0.75}Bi_{0.25}O_3}$ samples.  In the present investigation we have carried out dc magnetization measurements on K-doped Ba$_{0.9}$K$_x$BiO$_3$ samples with $x \approx 0.2$.''\cite{Mattheiss1988}  This investigation resulted in superconducting onset temperatures up to about 22~K as noted above.

Cava and coworkers subsequently refined the synthesis procedures and obtained a superconducting onset temperature of 29.8~K in a single-phase cubic perovskite structure sample of Ba$_{0.6}$K$_{0.4}$BiO$_3$, thus firmly confirming the source of the superconductivity as the compound with the cubic perovskite structure.\cite{Cava1988}  The onset temperature was determined by Cava et al.\ from a magnetization measurement of the Meissner effect upon cooling the sample in an applied magnetic field of 19~Oe.  We note that direct Meissner effect measurements such as this on other materials are often not informative or conclusive due to flux trapping effects that prevent the magnetic flux from escaping from the sample upon cooling below $T_{\rm c}$.

\begin{acknowledgments}
I am grateful to Arthur Sleight and Leonard Mattheiss for communicating to me historical aspects of their discoveries of superconductivity in the non-transition metal perovskites.  Work at the Ames Laboratory was supported by the Department of Energy-Basic Energy Sciences under Contract No.~DE-AC02-07CH11358.
\end{acknowledgments}

\end{document}